\documentclass[twocolumn,superscriptaddress,letter]{revtex4}%
\usepackage{amssymb}
\usepackage{color}
\usepackage{graphicx}
\usepackage{dcolumn}
\usepackage{bm}
\usepackage[header,title,page,titletoc]{appendix}
\usepackage{amsmath}
\usepackage{amsfonts}%
\providecommand{\U}[1]{\protect \rule{.1in}{.1in}}

\begin{document}
\title{State-dependent Topological Invariants and Anomalous Bulk-Boundary
Correspondence in non-Hermitian Topological Systems}
\author{Xiao-Ran Wang}
\affiliation{Center for Advanced Quantum Studies, Department of Physics, Beijing Normal
University, Beijing 100875, China}
\author{Cui-Xian Guo}
\affiliation{Center for Advanced Quantum Studies, Department of Physics, Beijing Normal
University, Beijing 100875, China}
\author{Su-Peng Kou}
\thanks{Corresponding author}
\email{spkou@bnu.edu.cn}
\affiliation{Center for Advanced Quantum Studies, Department of Physics, Beijing Normal
University, Beijing 100875, China}

\begin{abstract}
The breakdown of the bulk-boundary correspondence in non-Hermitian (NH)
topological systems is an open, controversial issue. In this paper, to resolve
this issue, we ask the following question: \emph{Can a (global) topological
invariant completely describe the topological properties of a NH system as its
Hermitian counterpart?} Our answer is \emph{no}. One cannot use a global
topological invariant (including non-Bloch topological invariant) to
accurately characterize the topological properties of the NH systems. Instead,
there exist a new type of topological invariants that are absence in its
Hermitian counterpart -- \emph{the state-dependent topological invariants}.
With the help of the state-dependent topological invariants, we develop a new
topological theory for NH topological system beyond the general knowledge for
usual Hermitian systems and obtain an exact formulation of the bulk-boundary
correspondence, including state-dependent phase diagram, state-dependent phase
transition and anomalous transport properties (spontaneous topological
current). Therefore, these results will help people to understand the exotic
topological properties of various non-Hermitian systems.

\end{abstract}

\pacs{11.30.Er, 75.10.Jm, 64.70.Tg, 03.65.-W}
\maketitle

Topological systems, including topological insulators and topological
superconductors have become the forefront of research in condensed matter
physics for many years\cite{Kane2010,Qi2011,Thouless,Qi,Haldane,Kane and
Mele,Ali,chi,ban}. These gapped topological system are always characterized by
certain (global) topological invariants and have\ intrinsic topological
properties that are robust and immunes to perturbations. For two quantum
phases with different topological invariants, one cannot deform the ground
states from one quantum phase to the other without closing the energy gap. On
the other hand, non-Hermitian (NH) topological systems\ have been intensively
studied in both
theory\cite{Rudner2009,Esaki2011,Hu2011,Liang2013,Zhu2014,Lee,Y.
Xu,Leykam2017,Shen2018,Xiong2018,Kawabata2018,Z. Gong,Yao20181,Yao20182,F. K.
Kunst2018,C. Yin,K. Kawabata2018,Alvarez,S. Chen2018,A. McDonald,H.-G.
Zirnstein,L. Jin2019,C. H. Lee 2019,Ueda2019,K. Kawabata2019,H. Y.Zhou, C. H.
Liu2019,L. Herviou,K. Yokomizo2019,B. Zhou2019,W.Yi2019,F. Song20191,F.
Song20192,X.W. Luo2019,N. Okuma2019,E. J. Bergholtz,J. Y. Lee2019,W. B.
Rui2019,H. Schomerus2020,K.-I. Imura2019,L. Herviou2019,P.-Y. Chang,X.-X.
Zhang2020,C. Fang,N. Okuma2020,S. Longhi2020,X.R. Wang,T. Yoshida,K.
Kawabata2020,C. Wang} and experiments\cite{Zeuner2015,S. Weimann2017,L.
Xiao2017,H. Zhou2018,M. A.Bandres2018,A. Cerjan2019,K. Wang2019,H. Zhao2019,M.
Brandenbourger2019,L. Xiao arXiv2020,T. Helbig}. The topological properties of
NH systems show quite different properties as their Hermitian counterparts.
Recently, within the generalization of Altland-Zirnbauer (AZ) theory, the
classification of NH systems with topological bands is characterized by
different symmetry-protected topological invariants\cite{Z. Gong,K.
Kawabata2019,H. Y.Zhou}.

An open issue is the breakdown the bulk-boundary correspondence (BBC) in NH
systems that has recently become a subject of active and controversial
discussion \cite{Lee,Xiong2018,Kawabata2018,Yao20181,Yao20182,F. K.
Kunst2018,C. Yin,L. Jin2019,L. Herviou2019}. Due to the existence of NH skin
effect, the conventional approach of predicting boundary states from bulk
topological invariants for periodic systems does not provide a conclusive
physical picture. According to Ref.\cite{Yao20181}, it was known that it is
non-Bloch topological invariant that characterizes the topological properties
of the NH topological systems. However, the non-Bloch topological invariants
cannot predict the existence of the (singular) defective edge state (an edge
state on the ends of an one-dimensional (1D) topological system with NH coalescence).

Hence, to complete solve the open issue of the breakdown the bulk-boundary
correspondence in NH systems, we develop a new theory for non-Hermitian
topological system by proposing the state-dependent topological invariants. We
point out that it is the state-dependent topological invariants rather than a
global state-independent topological invariant that characterize the
non-Hermitian topological phases. With the help of effective edge Hamiltonian,
we show spontaneous EP phenomenon together with topological Hermitian-NH
transition for a given edge state. In addition, due to the unbalance of the
state-dependent topological invariants for the edge states on chemical
potential there exists spontaneous topological current for 2D non-Hermitian
Chern insulator.

\textit{State-dependent topological invariants for 1D\ NH topological
insulator:} Firstly, we take 1D nonreciprocal Su-Schrieffer-Heeger (SSH) model
as an example to introduce the state-dependent topological invariants and
provide a new description of bulk-boundary correspondence for 1D NH
topological system.

The Bloch Hamiltonian for a nonreciprocal SSH model under periodic boundary
condition (PBC) is $H_{\mathrm{PBC}}=\left(  t_{1}+t_{2}\cos k\right)
\sigma_{x}+\left(  t_{2}\sin k+i\gamma \right)  \sigma_{y}+\varepsilon
\sigma_{z}$ where $t_{1}$ and $t_{2}$ describe the intra-cell and inter-cell
hopping strengths, respectively. $\varepsilon$ is the staggered potential and
$\gamma$ describes the unequal intra-cell hoppings. $\sigma_{i}$'s are the
Pauli matrices acting on the (\textrm{A} or \textrm{B}) sublattice subspace.
In this paper, we set $t_{2}=1$. It was known that due to the NH skin effect
the bulk spectrum of the system becomes that of a NH Hamiltonian
$H_{\mathrm{OBC}}$ with open boundary condition (OBC). As a result, the
effective bulk Hamiltonian turns into\cite{Yao20181,Yao20182} $H_{\mathrm{OBC}%
}(k)=(\bar{t}_{1}+\bar{t}_{2}\cos k)\sigma_{x}+(\bar{t}_{2}\sin k)\sigma
_{y}+\varepsilon \sigma_{z}$ where the effective hopping parameters become
$\bar{t}_{1}=\sqrt{(t_{1}-\gamma)(t_{1}+\gamma)},\quad$and $\bar{t}_{2}%
=t_{2}.$ Here, $\mathcal{\hat{S}}_{\mathrm{NHP}}$ is a similar-transformation,
i.e., $\left \vert \mathrm{\psi}(k)\right \rangle \rightarrow \left \vert
\mathrm{\bar{\psi}}(k)\right \rangle =\left \vert \mathrm{\psi}(k-iq_{0}%
)\right \rangle =\mathcal{\hat{S}}_{\mathrm{NHP}}\left \vert \mathrm{\psi
}(k)\right \rangle $ or $|n\rangle \rightarrow|\bar{n}\rangle=e^{-q_{0}%
(n-1)}|n\rangle$ ($n$ denotes the cell number) with $e^{q_{0}}=\sqrt
{\frac{t_{1}-\gamma}{t_{1}+\gamma}}.$

To completely characterize the edge states, we introduce the state-dependent
topological invariants $\{ \bar{v}_{L},$ $\bar{v}_{R}\}=\{ \bar{v}_{\xi},$
$\xi=L,R\}$ where $\bar{v}_{L}$ and $\bar{v}_{R}$ are topological invariants
for the edge states at left and right, respectively. $\{ \bar{v}_{L},$
$\bar{v}_{R}\}$ are combination of Bloch topological invariants from
$H_{\mathrm{PBC}}$ and non-Bloch topological invariants from $H_{\mathrm{OBC}%
}$, i.e.,
\begin{equation}
\bar{v}_{L}=\bar{w}\cdot v_{L},\text{ }\bar{v}_{R}=\bar{w}\cdot v_{R}%
\end{equation}
where $v_{L}=\frac{1}{2\pi}\int dk\cdot \partial_{k}\varphi_{+}$ and
$v_{R}=\frac{1}{2\pi}\int dk\cdot \partial_{k}\varphi_{-}$ are the Bloch
winding number that are defined from the Hamiltonian under PBC
$H_{\mathrm{PBC}}.$ $\varphi_{\pm}=$Arg$(h_{\pm})$ and $h_{\pm}$ is described
by $h_{\pm}=h_{x}\pm ih_{y}$ ($h_{x}=t_{1}+t_{2}\cos k$, $h_{y}=t_{2}\sin
k+i\gamma$); $\bar{w}=\frac{1}{2\pi}\int_{-\pi}^{\pi}\partial \bar{\phi}(k)dk$
is the non-Bloch topological invariant that is defined from the Hamiltonian
under OBC $H_{\mathrm{OBC}}$, where $\bar{\phi}(k)=\tan^{-1}(\bar{h}_{y}%
/\bar{h}_{x})$ and $\bar{h}_{x}=\bar{t}_{1}+\bar{t}_{2}\cos k$, $\bar{h}%
_{y}=\bar{t}_{2}\sin k$.

The state-dependent topological invariants $\{ \bar{v}_{\xi},$ $\xi=L,R\}$
become a complete description of BBC for 1D NH topological systems:
\emph{There exist }$\left \vert \bar{v}_{L}\right \vert $ \emph{edge states at
left end and }$\left \vert \bar{v}_{R}\right \vert $\emph{ edge states at right
end. }The state-dependent topological invariants are not applied to the case
of $\varepsilon=0.$ The non-Hermitian SSH model with $\varepsilon=0$ is very
special and unstable to arbitrary perturbation breaking chiral symmetry. We
will discuss the case of $\varepsilon=0$ in supplementary materials in detail.

As a result, for the 1D NH SSH model, we have a state-dependent phase diagram
with four phases (See Fig.1(a)): phase I, phase II, phase III, phase IV. There
exist two kinds of topological phase transitions: the state-independent
topological transition at $\left \vert \bar{t}_{1}\right \vert =\left \vert
\bar{t}_{2}\right \vert $ is characterized by the changing of non-Bloch
topological invariant $\bar{w}$ for $H_{\mathrm{OBC}}$ from a trivial phase
with $\bar{v}_{L}=\bar{v}_{R}=0$ (or $\bar{w}=0$) to topological phase with
$\bar{v}_{L}\neq$ or $\bar{v}_{R}\neq0$ (or $\bar{w}=1$); the other at
$t_{1}\pm \gamma=\pm1$ is state-dependent that is characterized by the changing
of Bloch topological invariant $v_{L}$ or $v_{R}$ for $H_{\mathrm{PBC}}$ from
a topological phase with the edge states ($\left \vert \bar{v}_{L}\cdot \bar
{v}_{R}\right \vert =1$, and $\bar{w}=1$) to another without them ($\bar{v}%
_{L}\cdot \bar{v}_{R}=0$, and $\bar{w}=1$).

\begin{figure}[ptb]
\includegraphics[clip,width=0.5\textwidth]{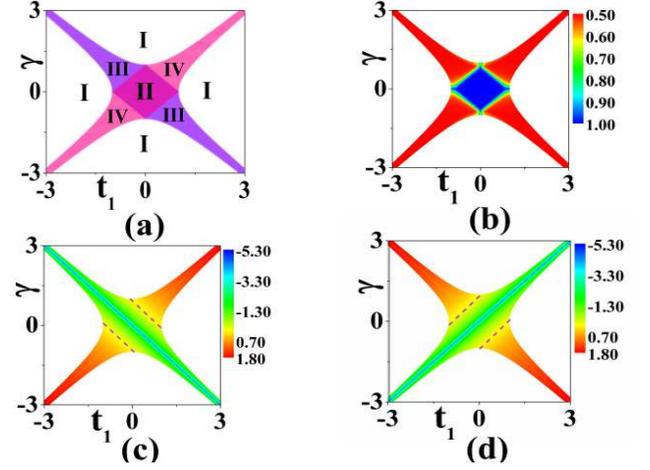}\caption{(Color online)
(a) State-dependent phase diagram: phase I with $\bar{v}_{L}=0$ and $\bar
{v}_{R}=0$ -- trivial phase without edge states; phase II with $\bar{v}_{L}=1$
and $\bar{v}_{R}=1$ -- topological phase with two edge states at left and
right ends; phase III with $\bar{v}_{L}=1$ and $\bar{v}_{R}=0$ -- topological
phase with only one edge state at left end; phase IV with $\bar{v}_{L}=0$ and
$\bar{v}_{R}=1$ -- topological phase with only one edge state at right end;
(b) The numerical results for BBC ratio $\gamma_{\mathrm{BBC}}=1-\frac
{|\langle \psi_{+}|\psi_{-}\rangle|}{2}$, for the case of $N=100,$
$\varepsilon=0.1$. In blue region we have a Hermitian phase. In red regions,
we have non-Hermitian phases with spontaneous EP phenomenon. Between the blue
region and red regions, topological Hermitian-NH transition occurs; (c) and
(d): The numerical results for off-diagonal term of the effective edge
Hamiltonian $\mathcal{\hat{H}}_{\mathrm{edge}},$ $\Delta^{+}$ (a) and
$\Delta^{-}$ (b). The dotted lines denote the topological
Hermitian--non-Hermitian transition with $\frac{\ln \bar{\Delta}^{\pm}}{N}=0$
shown in (a) and (b).}%
\end{figure}

To verify the validity of the state-dependent topological invariants $\{
\bar{v}_{\xi},$ $\xi=L,R\}$ and explore the corresponding topological
transitions for the 1D NH topological insulators, we write down the effective
Hamiltonian for the edge states $\mathcal{\hat{H}}_{\mathrm{edge}}=\left(
\begin{array}
[c]{cc}%
h_{11} & h_{12}\\
h_{21} & h_{22}%
\end{array}
\right)  $ where $h_{IJ}=\left \langle \mathrm{b}^{I}\right \vert \hat
{H}_{\mathrm{NH}}\left \vert \mathrm{b}^{J}\right \rangle ,$ $I,J=1,2$. $(%
\begin{array}
[c]{c}%
\left \vert \mathrm{b}^{1}\right \rangle \\
\left \vert \mathrm{b}^{2}\right \rangle
\end{array}
)$ are the basis under biorthogonal set of the edge states at left/right ends.
For the NH SSH model with $\gamma \neq0$, the effective edge Hamiltonian is
obtained as
\begin{equation}
\mathcal{\hat{H}}_{\mathrm{edge}}=\bar{\Delta}^{+}\tau^{+}+\bar{\Delta}%
^{-}\tau^{-}+\varepsilon \tau^{z}%
\end{equation}
where $\Delta^{+}=\bar{\Delta}e^{-Nq_{0}}=\frac{(t_{1}^{2}-t_{2}^{2}%
-\gamma^{2})}{t_{2}}(-\frac{t_{1}^{2}-\gamma^{2}}{t_{2}^{2}}e^{-2q_{0}}%
)^{N/2}$ and $\Delta^{-}=\bar{\Delta}e^{Nq_{0}}=\frac{(t_{1}^{2}-t_{2}%
^{2}-\gamma^{2})}{t_{2}}(-\frac{t_{1}^{2}-\gamma^{2}}{t_{2}^{2}}e^{2q_{0}%
})^{N/2}.$ $\bar{\Delta}$\ is the energy tunneling with the exponential decay
of number of unit cells $N.$ In thermodynamic limit $N\rightarrow \infty$,
although $\bar{\Delta}\rightarrow0,$ the results are non-trivial.

According to the off-diagonal term $\Delta^{+}=\bar{\Delta}e^{-Nq_{0}}$ or
$\Delta^{-}=\bar{\Delta}e^{Nq_{0}}$ of $\mathcal{\hat{H}}_{\mathrm{edge}}$,
there exists the competition between the exponential decay of $N$ from energy
tunneling $\bar{\Delta}\sim e^{\frac{N}{2}\ln(-\frac{t_{1}^{2}-\gamma^{2}%
}{t_{2}^{2}})}$ and the exponential increase with $N$ from NH similarity
transformation $e^{\pm Nq_{0}}$. Therefore, in thermodynamic limit (or
$N\rightarrow \infty$) there exist two phases: one is Hermitian phase with
$\left \vert \bar{\Delta}e^{\pm Nq_{0}}\right \vert \rightarrow0$, the other is
NH phase with $\left \vert \bar{\Delta}e^{\pm Nq_{0}}\right \vert \rightarrow
\infty$. In the Hermitian phase with $\left \vert \bar{\Delta}e^{\pm Nq_{0}%
}\right \vert \rightarrow0$, the effective edge Hamiltonian is reduced into
$\mathcal{\hat{H}}_{\mathrm{edge}}\rightarrow \varepsilon \cdot \tau^{z}$. Now,
the effect from the NH similarity transformation is irrelevant; On the other
hand, in the NH phase with $\left \vert \bar{\Delta}e^{\pm Nq_{0}}\right \vert
\rightarrow \infty$, the effective edge Hamiltonian is reduced into
$\mathcal{\hat{H}}_{\mathrm{edge}}\rightarrow \bar{\Delta}e^{-Nq_{0}}\tau^{+}$
or $\bar{\Delta}e^{Nq_{0}}\tau^{-}.$ Now, the effect from NH similarity
transformation dominates and becomes relevant. Although the total energy
splitting $E_{+}-E_{-}=2\sqrt{\varepsilon^{2}+\bar{\Delta}^{2}}$ is finite,
the system is at exceptional points (EPs) and we have singular defective edge
state with NH coalescence. For this reason, we call it \emph{spontaneous EP
phenomenon.}

At $\left \vert \bar{\Delta}e^{\pm q_{0}}\right \vert =1$, the topological
transition between Hermitian phase and NH\ phase with spontaneous EP occurs.
We call the state-dependent topological transition to be \emph{topological
Hermitian--NH transition}. Fig.1(c) and Fig.1(d) are the numerical results for
$\Delta^{+}=\bar{\Delta}e^{-Nq_{0}}$ or $\Delta^{-}=\bar{\Delta}e^{Nq_{0}},$
in which $\frac{\ln \Delta^{\pm}}{N}=0$ or $\left \vert \frac{t_{1}^{2}%
-\gamma^{2}}{t_{2}^{2}}e^{\pm2q_{0}}\right \vert =1$ denotes the topological
Hermitian--NH transition. This condition for topological Hermitian--NH
transition is just $t_{1}\pm \gamma=\pm1$ from phase with the edge states
($\left \vert \bar{v}_{L}\cdot \bar{v}_{R}\right \vert =1$, and $\bar{w}=1$) to
another without them ($\bar{v}_{L}\cdot \bar{v}_{R}=0$, and $\bar{w}=1$).\ To
verify this type of topological transitions and show the defectiveness of the
edge states, we calculate the BBC ratio, $\gamma_{\mathrm{BBC}}=1-\frac
{|\langle \psi_{+}|\psi_{-}\rangle|}{2}$, a quantity that characterizes number
anomaly of the edge states. If $\Upsilon_{\mathrm{BBC}}$ is $1$, there exists
the usual BBC with $\bar{v}_{L}=1$ and $\bar{v}_{R}=1$; if $\Upsilon
_{\mathrm{BBC}}$ is smaller than $1$, there exists defective edge states with
$\bar{v}_{L}=1$ and $\bar{v}_{R}=0$ or $\bar{v}_{L}=0$ and $\bar{v}_{R}=1.$ In
Fig.1(b), we show the numerical and analytic results that are consistent with
each other.

\textit{State-dependent topological invariants and spontaneous topological
current for 2D NH topological insulators}: Next, we consider a lattice model
of 2D NH Chern insulator - the 2D NH spin-orbital coupling model\cite{Qi}. The
Bloch Hamiltonian under PBC is $H_{\mathrm{PBC}}(k_{x},k_{y})=(\sin
k_{x})\sigma_{x}+(\sin k_{y}+i\gamma)\sigma_{y}+(m+\cos k_{x}+\cos
k_{y})\sigma_{z},$ where $\sigma_{x,y,z}$ are Pauli matrices. The NH
parameters $\gamma$ appear as \textquotedblleft imaginary Zeeman
field\textquotedblright. \ Suppose that the cylinder has periodic-boundary
condition along $x$-direction and open boundary condition along $y$-direction,
by doing similar transformation $\mathcal{\hat{S}}_{\mathrm{NHP}},$
$\left \vert \mathrm{\psi}(k)\right \rangle \rightarrow \left \vert \mathrm{\bar
{\psi}}(k)\right \rangle =\left \vert \mathrm{\psi}(k-iq_{0})\right \rangle
=\mathcal{\hat{S}}_{\mathrm{NHP}}\left \vert \mathrm{\psi}(k)\right \rangle $,
we have the non-Bloch \textquotedblleft cylinder Hamiltonian\textquotedblright%
, $H_{\mathrm{OBC,}y}(k_{x},\tilde{k}_{y})=H_{\mathrm{PBC}}(k_{x}\rightarrow
k_{x},k_{y}\rightarrow \tilde{k}_{y}-iq_{0},$ in which $e^{-2q_{0}(k_{x}%
)}=\frac{m+\cos k_{x}+\gamma}{m+\cos k_{x}-\gamma}$.

Then, we define the state-dependent topological invariants for edge states in
the 2D NH spin-orbital coupling model. For OBC along $y$-direction and PBC
along $x$-direction, the topological system exhibits modes localized on the
edges and the wave vectors $k_{x}=k=\frac{2\pi n}{L_{x}}$ ($n=1,2,...,N_{x}$)
are good quantum numbers. The state-dependent topological invariants are
defined as%
\begin{equation}
\{ \bar{v}_{\xi,k},k\in T^{1},\text{ }k\neq0,\  \xi=L,R\}
\end{equation}
where $\bar{v}_{k,L}=\mathcal{C}_{y}\cdot v_{k,L}$ are topological invariants
for all edge state at left end with wave vector $k$ and $\bar{v}%
_{k,R}=\mathcal{C}_{y}\cdot v_{k,R}$ are topological invariants for all edge
state at right end with wave vector $k$, respectively. Here, $\mathcal{C}%
_{y}=\frac{1}{2\pi i}\int dk_{x}d\tilde{k}_{y}\, \epsilon^{ij}\langle
\partial_{i}u_{L}(k_{x},\tilde{k}_{y})|\partial_{j}u_{R}(k_{x},\tilde{k}%
_{y})\rangle$ is the non-Bloch Chern number that is defined from the
Hamiltonian $H_{\mathrm{OBC,}y}(k_{x},\tilde{k}_{y})$ ($|u_{R}(k_{x},\tilde
{k}_{y})\rangle$ denotes the bulk state of biorthogonal set under OBC);
$v_{L,k}=\frac{1}{2\pi}\int dk_{y}\partial_{k_{y}}\varphi_{+}(k_{y},k)$ and
$v_{R,k}=\frac{1}{2\pi}\int dk_{y}\partial_{k_{y}}\varphi_{-}(k_{y},k)$ are
the Bloch winding number that are defined from the Hamiltonian
$H_{\mathrm{PBC}}(k_{x},k_{y})$. $h_{\pm}$ is described by $h_{\pm}=h_{x}\pm
ih_{y}$ ($h_{x}=(m+\cos k)+\cos k_{y}$, $h_{y}=\sin k_{y}+i\gamma$) and
$\varphi_{\pm}=$Arg$(h_{\pm})$.

Therefore, the state-dependent topological invariants $\{ \bar{v}_{\xi,k},$
$k\in T^{1},$ $k\neq0$,$\  \xi=L,R\}$ becomes a complete description of BBC for
2D NH topological systems: \emph{There exist }$\left \vert \bar{v}%
_{L,k}\right \vert $\emph{ edge states with wave vector }$\emph{k}$\emph{ at
left end and }$\left \vert \bar{v}_{R,k}\right \vert $\emph{ edge states with
wave vector }$\emph{k}$\emph{ at righ end. }The state-dependent topological
invariants are not applied to the case of $k=0$ and we will discuss the case
of $k=0$ in supplementary materials in detail. As a result, for the 2D NH
topological system with fixed parameter, we have a state-dependent phase
diagram via $k$ and $m$ (See Fig.2(a)).

\begin{figure}[ptb]
\includegraphics[clip,width=0.5\textwidth]{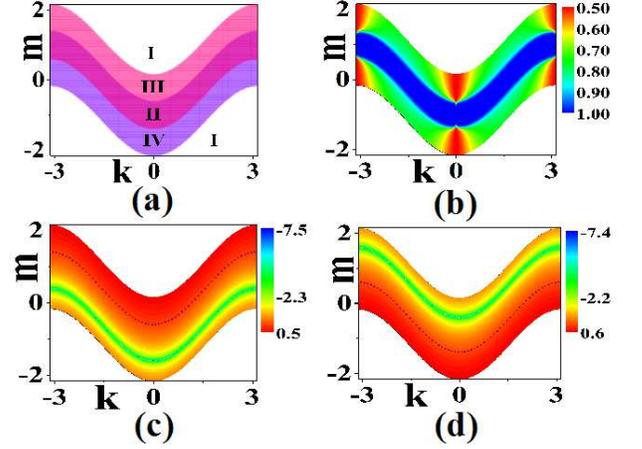}\caption{(Color online)
(a) is the state-dependent phase diagram via $m$ and $k$ with fixed
$\gamma=0.6$: phase I with $\bar{v}_{L,k}=0$ and $\bar{v}_{R,k}=0$ -- trivial
phase without edge states; phase II with $\bar{v}_{L,k}=1$ and $\bar{v}%
_{R,k}=1$ -- topological phase with two edge states at left and right ends;
phase III with $\bar{v}_{L,k}=1$ and $\bar{v}_{R,k}=0$ -- topological phase
with only one edge state at left end; phase IV with $\bar{v}_{L,k}=0$ and
$\bar{v}_{R,k}=1$ -- topological phase with only one edge state at right end;
(b) The numerical results for BBC ratio with fixed $\gamma=0.6$,
$\gamma_{k,\mathrm{BBC}}=1-\frac{|\langle \psi_{k,+}|\psi_{k,-}\rangle|}{2}$;
(c) and (d): The numerical results for off-diagonal term of the effective edge
Hamiltonian $\mathcal{\hat{H}}_{\mathrm{edge}}$ for the case of $\gamma=0.6$,
$\Delta_{k}^{+}$ (c) and $\Delta_{k}^{-}$ (d). The dotted lines correspond to
$\frac{\ln \bar{\Delta}_{k}^{\pm}}{N}=0,$ the topological
Hermitian--non-Hermitian transition shown in (a) and (b).}%
\end{figure}

To verify the validity of the state-dependent topological invariants, we
obtain the effective edge Hamiltonian for the NH spin-orbital coupling model.

Firstly, the effective edge Hamiltonian of edge states for Hermitian 2D Chern
insulator with $\gamma=0$ is obtained as $\mathcal{\hat{H}}_{\mathrm{edge}%
}=\tau^{z}\varepsilon_{k}+\tau^{x}\Delta_{k},$ where $\varepsilon_{k}=\sin k$
is the dispersion of the edge states of semi-infinite system and $\Delta
_{k}=((\cos k+m)^{2}-1)\cdot(\cos k+m)^{N_{y}}$ is tunneling strength. As a
result, the energy levels are $E_{k}=\pm \sqrt{(\sin k)^{2}+(\Delta_{k})^{2}}$.
In thermodynamic limit $N_{y}\rightarrow \infty$, $\Delta_{k}\rightarrow0,$ we
have $E_{k}\rightarrow \pm \sin k$ (or $E_{k}\rightarrow \sin k$ on left/right
edge and $E_{k}\rightarrow-\sin k$ on right/left edge). 

When considering the NH skin effect, we do an additional NH similarity
transformation $U_{\mathrm{edge}}=(%
\begin{array}
[c]{cc}%
1 & 0\\
0 & e^{-q_{0}(k)N_{y}}%
\end{array}
)$ on the effective edge Hamiltonian $\mathcal{\hat{H}}_{\mathrm{edge}}$,
i.e., $\tau^{x}\rightarrow U_{\mathrm{edge}}^{-1}\tau^{x}U_{\mathrm{edge}%
}=\tau^{x}\cosh(q_{0}(k)N_{y})-i\tau^{y}\sinh(q_{0}(k)N_{y}).$ As a result,
the effective edge Hamiltonian turns into
\begin{equation}
\mathcal{\hat{H}}_{\mathrm{edge}}=\bar{\Delta}_{k}^{+}\tau^{+}+\bar{\Delta
}_{k}^{-}\tau^{-}+\tau^{z}\sin k
\end{equation}
where$\bar{\Delta}_{k}^{+}=\bar{\Delta}_{k}e^{-N_{y}q_{0}(k)},$ $\bar{\Delta
}_{k}^{-}=\bar{\Delta}_{k}e^{N_{y}q_{0}(k)}$ and $\bar{\Delta}_{k}=((\cos
k+m)^{2}-1-\gamma^{2})\cdot(\gamma^{2}-(\cos k+m)^{2})^{N_{y}/2}.$ Under
$U_{\mathrm{edge}}$, the energy levels become $E_{\pm}(k)=\pm \sqrt{(\sin
k)^{2}+(\bar{\Delta}_{k})^{2}}.$

With the help of the effective edge Hamiltonian $\mathcal{\hat{H}%
}_{\mathrm{edge}}$, we show spontaneous EP phenomenon and topological
Hermitian-NH transition for a given edge state with wave vector $k$. At
$\frac{\ln \bar{\Delta}_{k}^{\pm}}{N}=0$ topological Hermitian-NH transition
occurs that is just the condition of $\left \vert ((\cos k+m)^{2}-\gamma
^{2})e^{\pm2q_{0}}\right \vert =1$ from phase with the edge states ($\left \vert
\bar{v}_{L,k}\cdot \bar{v}_{R,k}\right \vert =1$, and $\mathcal{C}_{y}=1$) to
another without them ($\bar{v}_{L,k}\cdot \bar{v}_{R,k}=0$, and $\mathcal{C}%
_{y}=1$). Fig.2(c) and Fig.2(d) are the numerical results for $\bar{\Delta
}_{k}^{+}$ or $\bar{\Delta}_{k}^{-},$ in which $\frac{\ln \bar{\Delta}_{k}%
^{\pm}}{N}=0$ denotes the critical points of the topological transitions. We
can also use $\Upsilon_{\mathrm{BBC}}$ to characterize the topological
properties of the edge states with wave vector $k,$ $\Upsilon_{k,\mathrm{BBC}%
}=1-|\langle \psi_{k,+}|\psi_{k,-}\rangle|/2.$ See the results in Fig.2(b).

In addition, we point out that there exists physics consequence of unbalance
of the state-dependent topological invariants for the edge states on chemical
potential -- \emph{the spontaneous topological current} for 2D NH Chern insulator.

For the system with open boundary condition (OPC), in general, the chemical
potentials at the ends of system may be different, i.e., $\mu_{L}$ and
$\mu_{R}$, respectively. We assume that the chemical potentials locate inside
the energy gap of the bulk states. So, the transport of the system mainly
comes from the edge states and we can apply the Landauer-Buttiker formalism on
the transport of edge states. According to the Landauer-Buttiker formalism,
the (Hall) current is defined as $I_{H}=-ev_{\mathrm{eff}}\cdot n.$ In
thermodynamic limit, the effective velocity of the charge carriers is
$v_{\mathrm{eff}}=\frac{1}{\hbar}\frac{\partial E_{k}}{\partial k}$ where
$E_{k}\sim \pm \sin k.$ The density of the charge carriers is
\begin{align}
n  &  =n_{L}-n_{R}=\mu_{L}\mathcal{D}(\mu_{L})-\mu_{R}\mathcal{D}(\mu_{R})\\
&  \sim(\mu_{L}\nu_{L,k_{L}}(E)-\mu_{R}\nu_{R,k_{R}}(E))D(E)\nonumber
\end{align}
\ where $\mathcal{D}(\mu_{L})=\nu_{L,k_{L}}(E)D(E)$ and $\mathcal{D}(\mu
_{R})=\nu_{R,k_{R}}(E)D(E).\  \nu_{L,k_{L}}(E)$ and $v_{R,k_{R}}(E)$ are the
state-dependent topological invariants for the edge states at left and right
sides, respectively. The wave vectors $k_{L}$ and $k_{R}$ are obtained by
calculating the following equations, $\sin k_{L}=\mu_{L}$ and $-\sin k_{R}%
=\mu_{R}$, respectively. As a result, considering $\mu_{L/R}=-eV_{L/R},$ we
derive the current for the NH 2D topological insulator,
\begin{align}
I_{H}  &  =\frac{e^{2}}{h}\mathcal{C}_{y}(V_{L}\nu_{L,k_{L}}(E)-V_{R}%
\nu_{R,k_{R}}(E))\nonumber \\
&  =\sigma_{0}(V_{L}\bar{\nu}_{L,k_{L}}(E)-V_{R}\bar{\nu}_{R,k_{R}}(E))
\end{align}
where $\sigma_{0}=\frac{e^{2}}{h}$ is unit of quantized Hall conductance.

In phase III and phase IV of Fig.2(a), when there doesn't exist an external
transverse electric field $\mu_{L}=\mu_{R}=\mu,$ due to the unbalance of the
state-dependent topological invariants for the edge states on chemical
potential, the electric current still exists, i.e., $I_{H}=\sigma_{0}\mu
(\bar{v}_{L,k_{L}}-\bar{v}_{R,k_{R}})\neq0.$ Because the current is
proportional to the unbalance of the state-dependent topological invariant
$(\bar{v}_{L,k_{L}}-\bar{v}_{R,k_{R}}),$ the spontaneous current is topological!

\textit{State-dependent Topological Invariants for NH }$d$-dimensional
\textit{topological insulators: }Finally, we generalize the theory of
state-dependent topological invariants to NH $d$-dimensional topological insulators.

A $d$-dimensional topological system can be approximatively described by
continuum model. Assuming PBCs in all directions, we consider the following
Hamiltonian for NH $d$-dimensional topological insulators as $H_{\mathrm{PBC}%
}(\mathbf{k})=\sum_{i=1}^{d}k_{i}\Gamma_{i}+(m-\frac{1}{2}\sum_{i=1}^{d}%
k_{i}^{2})\Gamma_{d+1}+i\gamma \Gamma_{j},$ where $\Gamma_{\mu}$ denote the
gamma matrices that satisfy $\{ \Gamma_{\mu},\Gamma_{\nu}\}=2\delta_{\mu \nu}$
and $m$ is the real mass parameter. $\gamma$ denotes the strength of the NH
term. With OBCs in the $j$-th direction and PBCs in all other directions, due
to the residue translation symmetry, $\tilde{\mathbf{k}}$ denotes the vector
of all momenta except $k_{j}$. Under a similar transformation $\mathcal{\hat
{S}}_{\mathrm{NHP}}=\mathrm{diag}\{1,\alpha,\cdots,\alpha^{N_{j}-1}\}
\otimes \left[  (1+\alpha)\mathbb{I}+i(1-\alpha)\Gamma_{j}\Gamma_{d+1}\right]
$, the Hamiltonian is transformed to $H_{\mathrm{OBC}}(\tilde{\mathbf{k}%
})=\sum_{i\neq j}k_{i}\Gamma_{i}+\sqrt{M_{k}^{2}-\gamma^{2}}\Gamma_{d+1},$
where $\alpha=\sqrt{(M_{k}-\gamma)/(M_{k}+\gamma)},$ $M_{k}=m-\frac{1}{2}%
\sum_{i\neq j}k_{i}^{2}$, and $N_{j}$ stands for the number of layers in the
$j$-th direction.

Then, to completely characterize the topological properties, we define the
state-dependent topological invariants for edge states in the $d$-D NH
topological model.

With OBCs in the $j$-th direction and PBCs in all other directions, due to the
residue translation symmetry, there exist edge states on left/right edges with
wave vector $\tilde{\mathbf{k}}$. The state-dependent topological invariants
for edge states are defined as
\begin{equation}
\{ \bar{v}_{\xi,\tilde{\mathbf{k}}},\text{ }\tilde{\mathbf{k}}\in
T^{d-1},\text{ }\tilde{\mathbf{k}}\neq0,\text{ and }\xi=L,R\}
\end{equation}
where $\bar{v}_{L,\tilde{\mathbf{k}}}=\mathcal{W}_{j}\cdot v_{L,\tilde
{\mathbf{k}}}$ for edge state at left end with wave vector $\tilde{\mathbf{k}%
}$ and $\bar{v}_{R,\tilde{\mathbf{k}}}=\mathcal{W}_{j}\cdot v_{R,\tilde
{\mathbf{k}}}$ for the edge state at right end with wave vector $\tilde
{\mathbf{k}}$. Here $\mathcal{W}_{j}$ is the d-dimensional non-Bloch
topological invariant that is defined from the Hamiltonian $H_{\mathrm{OBC}%
}(\tilde{\mathbf{k}})$. $v_{L,\tilde{\mathbf{k}}}$ and $v_{R,\tilde
{\mathbf{k}}}$ are the one-dimensional winding number that are defined from
the Hamiltonian $H_{\mathrm{PBC}}(\mathbf{k})$. $v_{L,\tilde{\mathbf{k}}%
}=\frac{1}{2\pi}\int dk_{j}\partial_{k_{j}}\varphi_{+}(k_{j},\tilde
{\mathbf{k}})$ and $v_{R,\tilde{\mathbf{k}}}=\frac{1}{2\pi}\int dk_{j}%
\partial_{k_{j}}\varphi_{-}(k_{j},\tilde{\mathbf{k}})$ are the Bloch winding
number that are defined from the Hamiltonian under closed boundary condition.
$h_{\pm}$ is described by $h_{\pm}=h_{x}\pm ih_{y}$ ($h_{x}=\sum_{i=1}%
^{d}\frac{1}{2}k_{i}^{2}$, $h_{y}=k_{j}+i\gamma$) and $\varphi_{\pm}%
=$Arg$(h_{\pm})$.\emph{ }The topological Hermitian--NH transition occurs at
$m-\sum_{i\neq j}^{d}\frac{1}{2}k_{i}^{2}=\pm \gamma \pm1.$

Therefore, the state-dependent topological invariants $\{ \bar{v}_{\xi
,\tilde{\mathbf{k}}},$ $\tilde{\mathbf{k}}\in T^{d-1},$ $\tilde{\mathbf{k}%
}\neq0,$ and $\xi=L,R\}$ become a complete description of BBC for $d$-D NH
topological systems: \emph{There exist }$\left \vert \bar{v}_{R,\tilde
{\mathbf{k}}}\right \vert $\emph{ edge states with wave vector }$\tilde
{\mathbf{k}}$\emph{ at left end and }$\left \vert \bar{v}_{L,\tilde{\mathbf{k}%
}}\right \vert $\emph{ edge states with wave vector }$\tilde{\mathbf{k}}$\emph{
at right end. }The state-dependent topological invariants are not applied to
the case of $\tilde{\mathbf{k}}=0$ and we will discuss the case of
$\tilde{\mathbf{k}}=0$ in supplementary materials in detail.

\textit{Conclusion and discussion}: In the end, we draw a brief conclusion.
The theory of state-dependent topological invariants for NH topological
insulators is developed. The key point of our theory is that each edge state
is characterized by state-dependent topological invariants $\{ \bar{v}%
_{\xi,\tilde{\mathbf{k}}},$ $\tilde{\mathbf{k}}\in T^{d-1},$ $\tilde
{\mathbf{k}}\neq0,$ and $\xi=L,R\}$ rather than a global (state-independent)
non-Bloch topological invariant $w_{j}$. To completely characterize the edge
states in a NH topological systems, one need to calculate all state-dependent
topological invariants. With the help of effective edge Hamiltonian
$\mathcal{\hat{H}}_{\mathrm{edge}}$, we derive the state-dependent phase
diagram and show spontaneous EP phenomenon together with (state-dependent)
topological Hermitian-NH transition for a given edge state with given wave
vector $\tilde{\mathbf{k}}$. In addition, there exists spontaneous topological
current $I_{H}$ for 2D NH Chern insulator that is proportional to the
unbalance of the state-dependent topological invariants $(\bar{v}_{L,k_{L}%
}-\bar{v}_{R,k_{R}})$ for the edge states on chemical potential $\mu_{L}%
=\mu_{R}=\mu$. In future, the theory of state-dependent topological invariants
can be applied to other types of topological systems, including topological
superconductors, higher order topological states, even the topological semi-metals.

\acknowledgments This work is supported by NSFC Grant No. 11674026, 11974053.
We thank Ya-Jie Wu, and Gao-Yong Sun\ for their helpful discussion.

\end{document}